\documentclass[12pt]{article}

\newcommand{\be}{\begin{equation}}
\newcommand{\ee}{\end{equation}}
\newcommand{\bi}[1]{\vspace{-3mm} \bibitem{#1}}

\usepackage{epsfig,amsmath,amssymb,graphics,graphicx} 

\begin{document}

\begin{center}
Theoretical and Mathematical Physics 158 (2009) 355-359
\end{center}

\begin{center}
{\Large \bf Fractional Integro-Differential Equations for}
\vskip 3 mm

{\Large \bf Electromagnetic Waves in Dielectric Media}
\vskip 5 mm

{\large \bf Vasily E. Tarasov }


\vskip 3mm

{\it Skobeltsyn Institute of Nuclear Physics, \\
Moscow State University, Moscow 119991, Russia \\ 
E-mail: tarasov@theory.sinp.msu.ru}

\end{center}

\begin{abstract}
We prove that the electromagnetic fields in dielectric media 
whose susceptibility follows a fractional power-law dependence 
in a wide frequency range can be described by differential equations  
with time derivatives of noninteger order. 
We obtain fractional integro-differential equations for
electromagnetic waves in a dielectric.
The electromagnetic fields in dielectrics 
demonstrate a fractional power-law relaxation.
The fractional integro-differential equations for electromagnetic waves 
are common to a wide class of dielectric media regardless of the type of physical 
structure, the chemical composition, or the nature of the polarizing species  
(dipoles, electrons, or ions).
\end{abstract}



\newpage
\section{Introduction} 

Debye formulated his theory of 
dipole relaxation in dielectrics in 1912 \cite{W1}.
A large number of dielectric relaxation measurements show that 
the classical Debye behavior is very rarely observed 
experimentally \cite{Jo2,Jrev,Ram}. 
Dielectric measurements by Jonscher for a wide class of various substances
confirm that different dielectric spectra 
are described by power laws \cite{Jo2,Jrev}.

For the majority of materials, the dielectric susceptibility 
in a wide frequency range follows a fractional power-law 
called the universal response \cite{Jo2,Jrev}. 
This law is found both in dipolar media beyond 
their loss peak frequency and in media where the polarization 
arises from movements of either ionic or electronic hopping charge carriers.
It was shown in \cite{Jo4} that 
the frequency dependence of the dielectric susceptibility
$\tilde \chi(\omega)=\chi^{\prime}(\omega)-i\chi^{\prime \prime}(\omega)$
follows a common universal pattern for virtually all kinds of media
over many decades of frequency, 
\be \label{W-3}
\chi^{\prime}(\omega) \sim \omega^{n-1} , \quad 
\chi^{\prime \prime}(\omega) \sim \omega^{n-1} , \quad 
(\omega \gg \omega_p) ,
\ee
and
\be  \label{W-4}
\chi^{\prime}(0)-\chi^{\prime}(\omega) \sim \omega^{m} , \quad 
\chi^{\prime \prime}(\omega) \sim \omega^{m} , \quad 
(\omega \ll \omega_p) ,
\ee
where $\chi^{\prime}(0)$ is the static polarization, $0 < n,m <1$, 
and $\omega_p$ is the loss peak frequency. 
We note that the ratio of the imaginary to the real component
of the susceptibility is independent of frequency. 
The frequency dependence given by equation (\ref{W-3}) 
implies that the imaginary and real components 
of the complex susceptibility at high frequencies 
satisfy the relation 
\be \label{W-5}
\frac{\chi^{\prime \prime}(\omega)}{\chi^{\prime }(\omega)} =
\cot \left( \frac{\pi n}{2} \right) , \quad  (\omega \gg \omega_p) .
\ee
Experimental behavior (\ref{W-4}) 
leads to a similar frequency-independent rule 
for the low-frequency polarization decrement, 
\be \label{W-6}
\frac{\chi^{\prime \prime}(\omega)}{ \chi^{\prime }(0)- \chi^{\prime }(\omega) }
=\tan \left( \frac{\pi m}{2} \right) , \quad (\omega \ll \omega_p) .
\ee

The laws of universal response for dielectric media \cite{Jo2,Jrev} 
can be described using fractional calculus \cite{SKM}.
The theory of integrals and derivatives of non-integer order goes back 
to Leibniz, Liouville, Riemann, Grunwald, and Letnikov \cite{SKM}. 
Fractional analysis has found many
applications in recent studies in mechanics and physics.
The interest in fractional equations has been growing continuously 
during the last few years because of numerous applications. 
In a short time, the list of applications has becomes long
(see, e.g., \cite{Zaslavsky2,Mainardi,Hilfer}). 
In Refs. \cite{NR,NP,Dielectric,First}, fractional calculus has used to explain
the nature of nonexponential relaxation, and equations were 
obtained containing operators of fractional integration and differentiation. 

Here, we prove that a fractional power-law frequency dependence 
in a time domain gives integro-differential equations 
with derivatives and integrals of noninteger order.
We obtain fractional equations that describe electromagnetic waves 
for a wide class of dielectric media.
The power laws of Jonscher are represented by 
fractional integro-differential equations. 
The electromagnetic fields in the dielectric media  
demonstrate universal fractional damping.
The suggested fractional equations are common (universal)
to a wide class of materials regardless of the type 
of physical structure, the chemical composition, or 
the nature of the polarizing species.


\section{Fractional equations for universal laws}

We consider Eqs. (\ref{W-3}) and (\ref{W-5}). 
For the region $\omega \gg \omega_p$, 
universal fractional power law (\ref{W-3})
can be presented in the form
\be \label{chi-1}
\tilde \chi(\omega)= \chi_{\alpha} \, (i \omega)^{-\alpha} , 
\quad (0<\alpha<1) 
\ee
with some positive constant $\chi_{\alpha}$ and $\alpha=1-n$. 
Here,
\[ (i\omega)^{\alpha}=|\omega|^{\alpha} 
\exp \{i \, \alpha \, \pi \, sgn(\omega)/2 \}. \]
It is easy to see that relation (\ref{W-5}) is satisfied for (\ref{chi-1}).

The polarization density ${\bf P}(t,r)$ can be written as
\be \label{New1} {\bf P}(t,r)={\cal F}^{-1} \left( \tilde {\bf P}(\omega,r) \right)=
\varepsilon_0 {\cal F}^{-1} \left(\tilde \chi(\omega)  \tilde {\bf E}(\omega,r) \right) 
, \ee
where $\tilde {\bf P}(\omega,r)$ is the Fourier transform ${\cal F}$ of ${\bf P}(t,r)$. 
Substitution of (\ref{chi-1}) into (\ref{New1}) gives
\[ {\bf P}(t,r)=\varepsilon_0 \chi_{\alpha} {\cal F}^{-1} 
\left( (i\omega)^{-\alpha} \tilde {\bf E}(\omega,r) \right) . \]

We note that the Fourier transform 
of the fractional Liouville integral \cite{SKM,KST}
\[ (I^{\alpha}_{+}f)(t)=\frac{1}{\Gamma(\alpha)} 
\int^{t}_{-\infty} \frac{f(t') dt'}{(t-t')^{1-\alpha}}  \]
is given by the relation 
(see Theorem 7.1 in \cite{SKM} and  Theorem 2.15 in \cite{KST}):
\[ ({\cal F} I^{\alpha}_{+}f)(\omega)=
\frac{1}{(i\omega)^{\alpha}} ({\cal F}f)(\omega) , \]
where $0<Re(\alpha)<1$ and $f(t) \in L_1(\mathbb{R})$, 
or $1\le p < 1/Re(\alpha)$ and $f(t) \in L_{p}(\mathbb{R})$.

Using the fractional Liouville integral and  
the fractional power law (\ref{chi-1}) for $\tilde\chi(\omega)$ 
in the frequency domain, we obtain
\be \label{Alpha}
{\bf P}(t,r)=\varepsilon_0 \chi_{\alpha}(I^{\alpha}_{+} {\bf E})(t,r) ,  
\quad (0<\alpha<1) . 
\ee
This equation shows that the polarization density ${\bf P}(t,r)$
in the high-frequency region is proportional to 
the fractional Liouville integral of the electric field ${\bf E}(t,r)$.


We consider Eqs. (\ref{W-4}) and (\ref{W-6}). 
For the region $\omega \ll \omega_p$, 
universal fractional power law (\ref{W-4}) can be presented as
\be \label{chi-2}
\tilde \chi(\omega)=\tilde \chi(0)-
\chi_{\beta} (i \omega)^{\beta} ,  \quad (0<\beta<1) 
\ee
with some positive constants $\chi_{\beta}$, 
$\tilde \chi(0)$, and $\beta=m$. 
It is easy to prove that equation (\ref{W-6}) is satisfied.

We note that the Fourier transforms of
the fractional Liouville derivative \cite{SKM,KST} 
\[ (D^{\beta}_{+}f)(t)=\frac{\partial^k}{\partial t^k}(I^{k-\beta}_{+}f)(t)
=\frac{1}{\Gamma(k-\beta)} \frac{\partial^k}{\partial t^k}
\int^{t}_{-\infty} \frac{f(t') dt'}{(t-t')^{\beta-k+1}} , \]
where $k-1 < \beta <k$, are given by the formula  
(see Theorem 7.1 in \cite{SKM} and  Theorem 2.15 in \cite{KST}):
\[ ({\cal F} D^{\beta}_{+}f)(\omega)=(i\omega)^{\beta} ({\cal F}f)(\omega) , \]
where $0<Re(\beta)<1$ and $f(t) \in L_1(\mathbb{R})$, 
or $1\le p < 1/Re(\beta)$ and $f(t) \in L_{p}(\mathbb{R})$.

Using the definition of the fractional Liouville derivative and  
fractional power laws (\ref{chi-2}), we can represent 
polarization density (\ref{New1}) in the form
\be \label{Beta}
{\bf P}(t,r)= \varepsilon_0 \tilde \chi (0) {\bf E}(t,r) 
- \varepsilon_0 \chi_{\beta}(D^{\beta}_{+} {\bf E})(t,r) ,  
\quad (0<\beta<1) . \ee
This equation shows that the polarization density ${\bf P}(t,r)$
in the low-frequency region is determined by 
the fractional Liouville derivative of the electric field ${\bf E}(t,r)$.

Relations (\ref{Alpha}) and (\ref{Beta}) can be considered universal laws. 
These equations with integro-differentiation of noninteger order 
allow obtaining fractional wave equations for electric and magnetic fields.


\section{Universal electromagnetic wave equation}

Here, we obtain fractional equations for electromagnetic fields 
in dielectric media.
Using the Maxwell equations, we obtain
\be \label{5}
\varepsilon_0 \frac{\partial^2 {\bf E}(t,r)}{\partial t^2}+
\frac{\partial^2 {\bf P}(t,r)}{\partial t^2}+
\frac{1}{\mu } \left( grad \, div {\bf E}- \nabla^2 {\bf E} \right) +
\frac{\partial {\bf j}(t,r)}{\partial t}=0 . 
\ee

For $\omega \gg \omega_p$, the polarization density ${\bf P}(t,r)$ 
is related with ${\bf E}(t,r)$ by equation (\ref{Alpha}). 
Substituting (\ref{Alpha}) in (\ref{5}), 
we obtain the fractional differential equation for the electric field
\be \label{EFE1}
\frac{1}{v^2} \frac{\partial^2 {\bf E}(t,r)}{\partial t^2}+
\frac{\chi_{\alpha}}{v^2} (D^{2-\alpha}_{+} {\bf E})(t,r) +
\left( grad \, div {\bf E}- \nabla^2 {\bf E} \right) =
-\mu\frac{\partial {\bf j}(t,r)}{\partial t} , 
\ee
where $0<\alpha<1$, and $v^2=1/(\varepsilon_0 \mu)$.
We note that $div {\bf E}\ne 0$ for $\rho(t,r)=0$.

For $\omega \ll \omega_p$, the fields ${\bf P}(t,r)$ and ${\bf E}(t,r)$ 
are related by Eq. (\ref{Beta}). 
In this case, equation (\ref{5}) becomes
\be \label{EFE2}
\frac{1}{v^2_{\beta}} \frac{\partial^2 {\bf E}}{\partial t^2}
-  \frac{a_{\beta}}{v^2_{\beta}}  (D^{2+\beta}_{+} {\bf E})+ 
\left( grad \, div {\bf E}- \nabla^2 {\bf E} \right) =
-\mu\frac{\partial {\bf j}}{\partial t} , \quad (0<\beta<1) ,
\ee
where
\[ v^{2}_{\beta}=\frac{1}{ \varepsilon_0 \mu \, [1+  \tilde \chi (0)] } , \quad
a_{\beta}= \frac{\chi_{\beta}}{1+  \tilde \chi (0) } . \]
Equations (\ref{EFE1}) and (\ref{EFE2}) describe the time evolution of 
the electric field in dielectric media.
These equations are fractional differential equations \cite{KST}
with derivatives of the orders $2-\alpha$ and $2+\beta$.


Using the Maxwell equations, we obtain the equation for the magnetic field
\be  \label{mag5}
\frac{\partial^2 {\bf B}(t,r)}{\partial t^2}= 
\frac{1}{\varepsilon_0  \mu} \nabla^2 {\bf B}(t,r)+
\frac{1}{\varepsilon_0} \frac{\partial }{\partial t} curl {\bf P}(t,r)+
\frac{1}{\varepsilon_0} curl {\bf j}(t,r). 
\ee
In experiments, the field ${\bf B}(t,r)$ can be presented as 
${\bf B}(t,r)=0$ for $t \le 0$, and ${\bf B}(t,r) \ne 0$ for $t>0$. 
For $\omega \gg \omega_p$, 
the polarization density ${\bf P}(t,r)$ is related to ${\bf E}(t,r)$
by equation (\ref{Alpha}), which leads to 
the fractional differential equation for magnetic field in the form
\be \label{MFE1}
\frac{1}{v^2}\frac{\partial^2 {\bf B}(t,r)}{\partial t^2}+
\frac{\chi_{\alpha}}{v^2} \left( _0D^{2-\alpha}_{t} {\bf B} \right)(t,r)
- \nabla^2 {\bf B}(t,r) =\mu \, curl \, {\bf j}(t,r) , 
\ee
where $0<\alpha<1$, $v^2=1/(\varepsilon_0  \mu)$, and $_0D^{2-\alpha}_{t}$
is the Riemann-Liouville derivative \cite{KST} on $[0,\infty)$ such that
\[ ( _0D^{2-\alpha}_{t}f)(t)=
\frac{1}{\Gamma(\alpha)} \frac{\partial^2}{\partial t^2}
\int^{t}_{0} \frac{f(t') dt'}{(t-t')^{1-\alpha}} , 
\quad (0 < \alpha <1) . \]

For $\omega \ll \omega_p$, we obtain
\be  \label{MFE2}
\frac{1}{v^2_{\beta}} \frac{\partial^2 {\bf B}(t,r)}{\partial t^2}-
\frac{a_{\beta}}{v^2_{\beta}} \left( _0D^{2+\beta}_{t} {\bf B} \right)(t,r)
- \nabla^2 {\bf B}(t,r) =\mu \, curl \, {\bf j}(t,r) , 
\ee
where $0<\beta<1$, and
\[ v^2_{\beta} =\frac{1}{\varepsilon_0 \mu \, [1+ \tilde \chi (0) ]} , \quad
a_{\beta}=\frac{\chi_{\beta} }{1+ \tilde \chi (0)} . \]
Equations (\ref{MFE1}) and (\ref{MFE2}) are
fractional differential equations that 
describe the magnetic field in dielectric media and 
demonstrate a power-law relaxation. 
They can be written in a general form.
Such a general fractional differential equation for 
the magnetic field has the form
\be \label{MM}
( _0D^{\alpha}_t {\bf B})(t,r)-
\lambda_1  \left( _0D^{\beta}_{t} {\bf B} \right)(t,r)
- \lambda_2 \nabla^2 {\bf B}(t,r) ={\bf f}(t,r), \ee
where $1\le\beta<\alpha<3$.
The curl of the current density of free charges 
is regarded as an external source:  
${\bf f}(t,r)= \mu \lambda_2 \, curl \, {\bf j}(t,r)$. 
Equation (\ref{MM}) yields Eq. (\ref{MFE1}) for $\alpha=2$, $1<\beta<2$, and
\[ \lambda_1= - \chi_{\alpha}, \quad \lambda_2=v^2=1/(\varepsilon_0  \mu) . \]
Equation (\ref{MFE2}) can be written in form (\ref{MM}) 
for $2<\alpha<3$, $\beta=2$, and
\[ \lambda_1= \frac{1}{a_{\beta}}=
\frac{1+\tilde \chi(0)}{\chi_{\beta}} , \quad 
\lambda_2=-\frac{v^2_{\beta}}{a_{\beta}}=
\frac{-1}{\varepsilon_0\mu \chi_{\beta}} . \]
An exact solution of Eq. (\ref{MM}) can be written in terms of 
Wright functions \cite{KST} using Theorem 5.5 in \cite{KST}.
We note that Wright functions can be represented as derivatives 
of the Mittag-Leffler function $E_{\alpha,\beta}[z]$ (see \cite{KST}).
Solutions of equation (\ref{MM}) describe the fractional power-law damping of 
the magnetic field in dielectric media.
An important property of the evolution described by the fractional 
differential equations is that the solutions have fractional power-law tails.

\section{Conclusion}

We have prove that
the electromagnetic fields and waves in a wide class of 
dielectric media must be described by 
fractional differential equations with derivatives of the order 
$2-\alpha$ and $2+\beta$, where $0<\alpha<1$ and $0<\beta<1$.
The parameters $\alpha=1-n$ and $\beta=m$ are defined by 
the exponents $n$ and $m$ in the experimentally measured 
frequency dependence of the dielectric susceptibility,  
called the universal response laws. 
An important property of the dynamics described by fractional 
differential equations for electromagnetic fields
is that the solutions have fractional power-law tails.
The suggested fractional integro-differential equations 
for the universal electromagnetic waves in dielectrics  
are common (universal) to a wide class of media 
regardless of the type of physical structure, the chemical composition,
or the nature of the polarizing species  
(dipoles, electrons, or ions).

We note that the differential equations 
with derivatives of noninteger order proposed for
describing the electromagnetic field in dielectric media 
can be solved numerically.
For example, the Grunwald-Letnikov discretization scheme \cite{SKM} 
is used for numerically model the electromagnetic field in dielectrics
described by fractional differential equations. 
For small fractionality of $\alpha$ (or $\beta$), 
an $\varepsilon$-expansion \cite{TZ2} in the small parameter 
$\varepsilon=\alpha$ (or $\varepsilon=1-\beta$) can be used. 
We note that a possible physical interpretation of 
fractional integrals and derivatives can be connected with memory effects
or fractal properties of media (see, e.g., \cite{Nig,Stan}).


\end{document}